# Data-driven Prediction of Ionic Conductivity in Solid-State Electrolytes with Machine Learning and Large Language Models

Haewon Kim[1,#], Taekgi Lee[1,#], Seongeun Hong[1], Kyeong-Ho Kim[2], and Yongchul G. Chung[1,3,a]

[1]School of Chemical Engineering, Pusan National University, Busan, Korea (South) 46241

[2]Department of Materials Science and Engineering, Pukyong National University, Busan, 48513, Korea (South) 48513

[3]Graduate School of Data Science, Pusan National University, Busan, Korea (South) 46241

[#]These authors contributed equally.

[a]Author to whom correspondence should be addressed: drygchung@gmail.com

**TOPICS:** Machine learning, Large-language models, Solid-state batteries, Ionic conductivity

**Abstract**

Solid-state electrolytes (SSEs) are attractive for next-generation lithium-ion batteries due to improved safety and stability, but their low room-temperature ionic conductivity hinders practical application. Experimental synthesis and testing of new SSEs remain time-consuming and resource-intensive. Machine learning (ML) offers an accelerated route for SSE discovery; however, composition-only models neglect structural factors important for ion transport, while graph neural networks (GNNs) are challenged by the scarcity of structure-labeled conductivity data and the prevalence of crystallographic disorder in CIFs. Here, we train two complementary predictors on the same room-temperature, structure-labeled dataset (n = 499). A gradient-boosted tree regressor (GBR) model using stoichiometric descriptors alone achieves a test MAE of 1.108 in log(S/cm); adding geometric descriptors (combined MAE = 1.172) does not lower the test error but reveals complementary structural information through Shapley Additive exPlanations (SHAP), which shows that stoichiometric descriptors, particularly the oxygen ratio, dominate feature importance (seven of the top ten features), with three geometric descriptors (density, $L_{max}$, $L_{min}$) also contributing meaningfully. In parallel, we fine-tune large language models (LLMs) using compact text prompts derived from CIF metadata (formula with optional symmetry and disorder tags), avoiding direct use of raw atomic coordinates. Notably, while Mistral-7B achieves the lowest absolute error (MAE = 0.798 in log(S/cm)), Qwen3-8B demonstrates the best overall ranking performance (SRCC = 0.849) using formula and disorder information, eliminating the need for numerical feature extraction from CIF files. Together, these results show that global geometric descriptors improve tree-based predictions and enable interpretable structure-property analysis, while LLMs provide a competitive low-preprocessing alternative for rapid SSE screening.

# I. INTRODUCTION

Electrolytes are essential components of rechargeable batteries because they enable ionic transport between the cathode and anode and mediate charge transfer at the electrode-electrolyte interfaces. Solid-state electrolytes (SSEs) have emerged as critical components for next-generation batteries, offering improved thermal stability, nonvolatility, and greatly reduced flammability compared to conventional liquid electrolytes[1]. Beyond these intrinsic safety advantages, SSEs also open the possibility of achieving higher energy densities by enabling the integration of Li-metal anodes and high-voltage cathodes, which are challenging to realize with liquid electrolytes[2]. Achieving room-temperature ionic conductivities approaching those of liquid electrolyte ($\sim 10^{-3}$-$10^{-2}$ S/cm) remains a central materials challenge for practical solid-state batteries[3]. A wide range of SSE chemistries have been explored in recent years, such as $Li_{10}GeP_2S_{12}$ (LGPS) and lithium argyrodites (e.g., $Li_6PS_5X$ where X = Cl, Br, I), and demonstrate high ionic conductivity, often exceeding $10^{-3}$ S/cm; however, they generally suffer from limited oxidative stability and interfacial reactivity[4]. In contrast, oxide-based SSEs, such as garnet-type (e.g., LLZO), NASICON-type (e.g., LATP), and perovskite-type (e.g., LLTO) offer a comparatively wide voltage window and greater stability but often fall short in ionic conductivity ($10^{-7}$-$10^{-3}$ S/cm)[5]. Such relatively low ionic conductivity at room temperature remains a significant bottleneck to the widespread adoption of SSEs[6].

Experimental synthesis and testing of solid-state electrolyte materials in search of commercially viable ionic conductivities are both resource-intensive and time-consuming. To accelerate materials discovery, computational methods have recently been applied to estimate ionic transport properties in solid-state electrolytes. However, these computational approaches also present additional challenges. Classical molecular dynamics (MD) can estimate diffusion and conductivity (e.g., through mean-squared displacement combined with the Nernst-Einstein relation or via Green-Kubo formalism[7]), but accuracy is highly dependent on the quality and transferability of the employed interatomic potentials, particularly their ability to reproduce migration barriers and defect energetics that govern ion transport[8, 9]. Alternatively, ab-initio molecular dynamics (AIMD) based on density functional theory (DFT) can improve the description of ion motion by using a first-principles potential energy surface. However, the generation of AIMD trajectories across a large number of materials is computationally intensive, requiring considerable time and resources[10]. While the Nernst–Einstein relation is commonly used to estimate conductivity from diffusivities, it neglects correlation effects (collective ion motion) that can be significant in highly conducting SSEs[11]. Additionally, AIMD simulations are performed at high temperatures to accelerate diffusion, and the room temperature ionic conductivity is estimated by extrapolating from these data. Such extrapolations can be unreliable for systems that exhibit non-linear Arrhenius behavior[12]. To overcome the limited timescales accessible by AIMD, recent efforts have focused on developing and applying machine-learning interatomic potentials (MLP) to generate trajectories on the nanosecond- to microsecond-scale. In selected systems, MLP-based MD simulations have shown encouraging agreement with experiment[12-14].

Recent advances in machine learning (ML) and artificial intelligence (AI) have enabled direct prediction of ionic conductivity across wide composition space, provided that high-quality data sets are available. Hargreaves et al.[15] released the Liverpool Ionics Data set (LiIon, 2023), which compiles 820 experimentally reported composition-conductivity pairs across 5-873 °C. Using this data set, they demonstrated composition-based machine learning (ML) screening, including attention-based architecture such as CrabNet[16]. Subsequent studies expanded these compilations and evaluated a variety of composition-based regression[17,18,19]. Although ML algorithms have improved composition-based models, their performance remains constrained by reliance on composition alone. Since ionic conductivity in solid-state electrolytes arises from ions hopping between adjacent vacant sites, incorporating crystal structure information and defect/disorder-related descriptors (e.g., site occupancies and vacancy distributions) in training may further enhance the predictive accuracy of these models.

Therrien et al.[20] recently introduced the Open solid Battery Electrolytes with Li: an eXperimental dataset (OBELiX, 2026), which integrates both compositional and structural information to benchmark the role of structure in model training. OBELiX contains 599 experimentally measured room-temperature ionic conductivities with corresponding compositions, drawn from the LiIon data set and the Laskowski data set[21]. Of these, 321 crystal structures (CIFs) were obtained from the Inorganic Crystal Structure Database (ICSD), the Materials Project, and manual collection of data in the literature. The composition data were used to train conventional ML models (Random Forest, Multi-Layer Perceptron), while CIF data were used to benchmark graph neural network (GNN) models. In their benchmark, composition-based models outperformed several crystal-graph neural network (GNN) models trained on the smaller CIF subset. They attributed this to two limitations of the structural data set: (i) the small-data regime and (ii) structural disorder in crystallographic data. Such disorder, arising from partially occupied or shared atomic sites that average multiple local configurations into a single structure, directly affects the local ionic environment[22, 23]. In practice, partial occupancy and vacancy information are difficult to represent in structure-only graph encoding, and in small-data regimes these factors can degrade model performance. These observations motivate the exploration of alternative ML architectures that can jointly encode compositional and structural information in representations more amenable to learning complex structure-property relationships.

As an alternative approach to incorporate structural information as features, recent studies have explored the use of large language models (LLMs) to process materials data in interpretable textual forms. LLMs can capture detailed structural features of crystalline materials, including disorder-related information, through textual representations. LLMs have been applied widely to predict properties of diverse systems, including zeolites[24], MOFs[25], transition metal complexes[26], drugs[27], alloys[28, 29] and polymers[30]. Notably, Rubungo et al.[31] demonstrated that LLMs can predict crystalline materials properties by fine-tuning on text-based descriptions derived from CIFs using tools such as Robocrystallographer[32].

These developments suggest that combining structural descriptors with modern ML architecture could overcome the limitations of composition-only models. However, the scarcity of experimentally determined structures for many SSE compositions remains a barrier. In this work, we develop two complementary data-driven predictors for room-temperature ionic conductivity that differ only in how crystal structures are represented. First, we curate a structure-labeled dataset of 499 inorganic SSEs by combining experimentally reported CIFs (OBELiX) with 152 additional USPEX-generated structures relaxed using CHGNet, plus a small set of DFT-optimized structures from MP/OQMD. Next, we train a gradient-boosted tree regressor (GBR) on engineered stoichiometric and geometric descriptors and interpret the learned structure-property relationships using SHAP to identify which global features (e.g., lattice scale and accessible volume) most strongly correlate with conductivity. In parallel, we fine-tune multiple large language models (LLMs) using compact text prompts constructed from CIF metadata (formula and selected crystallographic metadata, when available) to evaluate whether competitive accuracy can be obtained without direct parsing of raw CIF coordinates. Because both approaches are trained and evaluated on the same data split and target definition, the results provide a controlled comparison between numerical feature engineering and text-based learning for SSE screening.

## II. Methods

### A. Dataset Curation

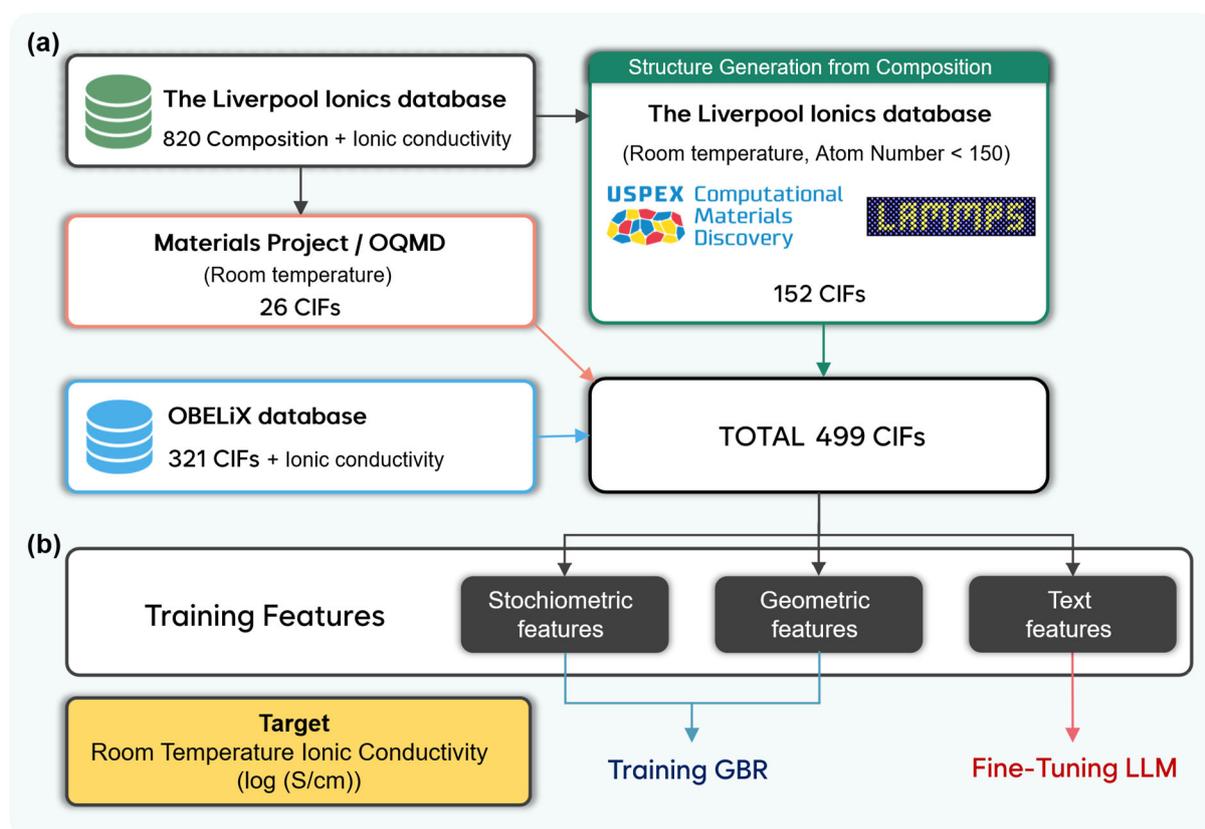

**Figure 1.** (a) Data curation and (b) model training part of overall workflow.

The overall dataset curation and model training workflow of this study is summarized in **Figure 1**. The dataset was constructed by combining composition data collected from the LiIon database with CIF structures from the OBELiX dataset and our generated dataset. In addition, DFT optimized structure files were collected from the materials project (MP) and the open quantum materials database (OQMD) based on the LiIon database. In this process, structures with stoichiometric errors resulting in improper charge balance in the original dataset were excluded. The schematic of the dataset construction is shown in **Figure 1(a)**. Based on this dataset, features were generated, and predictive models were trained as shown in **Figure 1(b)**. Details of each step are discussed in the following sections.

## B. Structure Generation and Validation

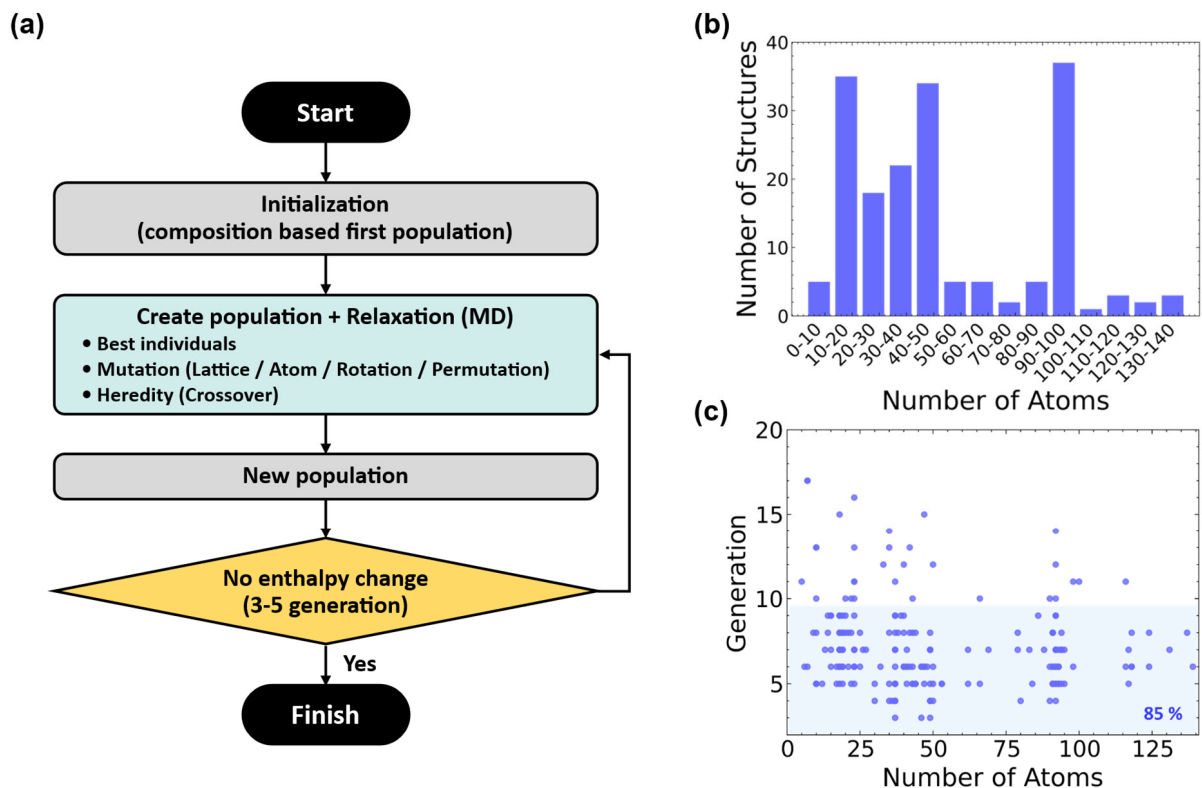

**Figure 2.** (a) Workflow of structure generation in USPEX based on a genetic algorithm. (b) Distribution of atom numbers in the generated structures (c) Number of generations as a function of atom numbers.

We employed the USPEX (Universal Structure Predictor: Evolutionary Xtallography)[33-35] software to generate crystal structures from composition data of LiIon database. The overall structure generation algorithm of USPEX is shown in **Figure 2(a).** USPEX generates structures by constructing a random initial population from the stoichiometric formula, followed by energy minimization. In genetic algorithm-based structure generation, thermodynamically stable structures are identified through iterative optimization involving heredity, mutation, and permutation. For variation operators in USPEX, atom mutation involves slight displacements of atomic positions, while permutation swaps the chemical identities of randomly chosen atom pairs. The heredity operator combines spatially coherent slabs from two parent structures, with their lattice vector matrices serving as weighted averages[36]. Although computationally intensive, these generation methods reduce the risk of trapping in local minima of the potential energy surface. At each generation, enthalpies of the candidate structures are calculated after energy minimization, and the structures with the lowest enthalpy values are selected to generate the next population. To increase computational efficiency of the search, we replaced density functional theory (DFT) calculation of enthalpy evaluation with all-atom machine learning potential (MLP)-based molecular dynamics (MD) simulations. To initiate the generation process, all chemical formulas were reduced to their simplest integer stoichiometric form, and the number of atoms per structure was restricted to fewer than 150 to ensure computational feasibility. **Figure 2(b)** shows the distribution of atom counts in the structures generated from reduced compositions. In each generation, structural relaxation was carried out until the enthalpy values converged within the same range over 3–5 successive generations. Approximately 85% of the structures achieved enthalpy convergence before the 10th generation. The number of generations as a function of atom number is shown in **Figure 2(c).**

MD simulations were carried out using the LAMMPS (2 Aug 2023 release)[37] using periodic boundaries in all directions. The CHGNet[38] model is a machine-learned interatomic potential trained on energies, forces, stresses, and magnetic moments from the Materials Project trajectory data set. CHGNET has been shown to perform reliably even for systems containing transition metals[39]. Benchmarking studies have also demonstrated that CHGNet reproduces AIMD results with near-DFT accuracy across a broad range of lithium-ion solid electrolytes[38]. In particular, CHGNet accurately captures highly nonlinear diffusion phenomena such as the sharp reduction in activation energy and the emergence of activated Li diffusion networks in garnet-type materials in excellent agreement with DFT benchmarks. These results indicate that CHGNet reliably models the strong local Li-Li interactions and complex diffusion environments characteristic of solid-state ionic conductors. We used CHGNET-LAMMPS-0.3.0 pair coefficient to evaluate energies and forces of generated structures. For both atomic and cell relaxation, the Nosé-Hoover thermostat and barostat were used as implemented in the LAMMPS. The system was equilibrated by NPT ensemble at 50 K and 1000 bar, with thermostat and barostat damping constants of 1 ps. The neighbor list was updated every step with a cutoff distance of 8 Å. Each simulation was run for 1,000 steps with 1 fs timestep (total of 1 ps) for integrating Newton's equations of motion. Properties($E_{pot}$, $E_{kin}$, $E_{tot}$, $T, p, V$) are recorded every step and atomic configurations saved every 50 steps for subsequent analysis. In total, 152 structures were generated using the USPEX method and the final MD relaxations using MLP model were conducted under the same conditions (NPT ensemble; 50 K and 1000 bar) as the enthalpy relaxation for an additional 20 ps simulation for all structures.

**C. Feature Extraction and Models**

To develop the ML models, we extracted features from CIFs obtained from the snapshot at the end of 20 ps relaxation simulations and the original CIFs from the OBELiX, Materials Project (MP), and Open Quantum Materials Database (OQMD). The ionic conductivity data corresponding to the structures were collected from the LiIon and OBELiX datasets. Stoichiometric features were employed to reflect the compositional properties of the structures, and geometric features from CIFs were generated to capture the structural characteristics of the SSEs. In the case of LLMs, text-based properties that integrate both stoichiometric and geometric features are required. Therefore, text-based structural information from CIF files was utilized for LLM fine-tuning. All types of features used for model training are presented in **Figure 3**.

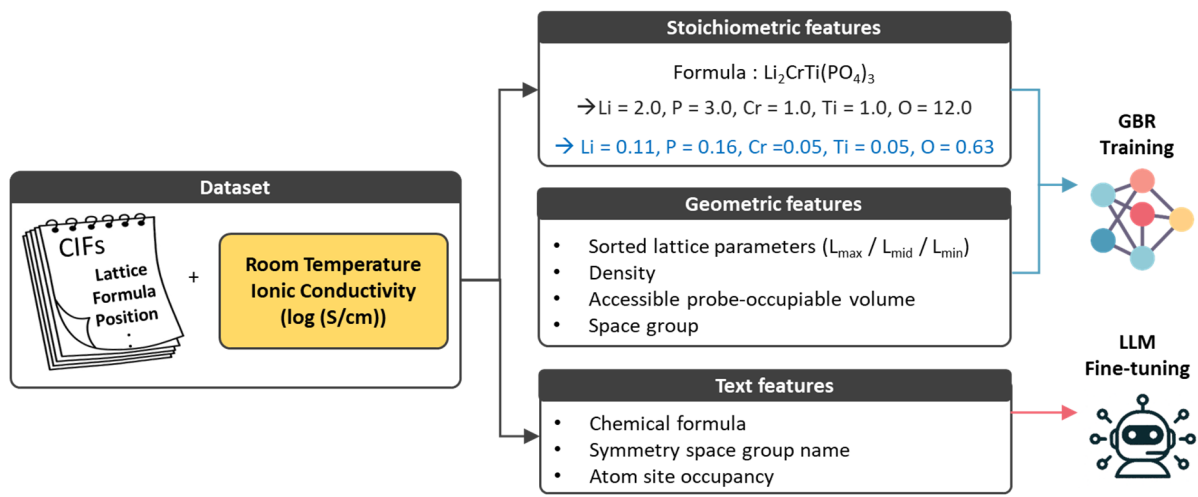

**Figure 3.** Features generated for model training. Features are extracted from the collected CIFs and Stoichiometric, geometric and text-based features were generated for training models.

**Gradient Boosting Regression (GBR) Training**

For the regression model, we employed the XGBOOST GBR algorithm to develop a predictive model for experimental ionic conductivity. For the stoichiometric features, the compositions were obtained using the chemparse[40] Python package. The elemental ratios contained in each composition were used as stoichiometric features. Geometric features are generated from original CIFs. To avoid axis-labeling bias arising from arbitrary crystallographic conventions, the three lattice parameters (a, b, c) were sorted by magnitude and relabeled as $L_{max}$, $L_{mid}$, and $L_{min}$ rather than being used directly. Vacancy volume-related descriptors were computed based on analysis of the Voronoi network, implemented in the Zeo++3.0 software[41-43]. Ion transport mechanism for solid electrolytes usually involves the movement of ions originating from a stable site via high energy transition state to another stable site[44-46]. The movement of ions is expected to be affected by lattice density and the porosity, which were regarded as key factors in ionic conductivity[47]. Accordingly, we used density (g/cm$^3$), gravimetric accessible probe-occupiable volume (POAV, cm$^3$/g), the reordered lattice parameters ($L_{max}$, $L_{mid}$, $L_{min}$), and space group as the geometric features. All features were processed through an integrated code to produce .csv files from CIFs.

Prior to training, we standardized all input features using z-score normalization implemented with StandardScaler in scikit-learn[48] (removing the mean and scaling to unit variance). The dataset was partitioned into a training set (80%, n=396) and a testing set (20%, n=103) using a Monte Carlo Group Split strategy, following the leakage-prevention principle adopted in the OBELiX benchmark[20]. To prevent data leakage, we enforced a strict grouping policy based on chemical composition; all entries with the same stoichiometric formula were assigned exclusively to either the training or testing set. This approach ensures that the model is evaluated on its ability to generalize to novel chemical spaces rather than memorizing redundant measurements of known materials.

The split was optimized through an iterative stochastic process (n = 200 iterations). In each iteration, the statistical similarity between the target distributions ($\log_{10} \sigma$) of the training and testing sets was evaluated using the two-sample Kolmogorov-Smirnov (K-S) test in SciPy[49]. The final split was selected by minimizing the K-S statistic ($D_{n,m}$), achieving a highly balanced partition with a minimal K-S value of 0.0510. This rigorous splitting method results in consistent target statistics (Training: $\mu = -5.736$, $\sigma = 2.819$; Testing: $\mu = -5.795$, $\sigma = 2.882$), providing an unbiased benchmark for the predictive performance of the GBR model. During training, 6-fold cross validation was implemented using the scikit-learn[48] library and the hyperparameter tuning was performed using the scikit-optimize library. Hyperparameters were tuned using Bayesian optimization, with the goal of minimizing the mean squared error (MSE) of the model. Optimized parameters are summarized in **Table 1**.

**Table 1.** Parameters of GBR models for ionic conductivity prediction.

| Model | Parameters | Values |
|---|---|---|
| Gradient Boosting Regression (GBR) | n_estimators | 29 |
| | reg_lambda | 30.991 |
| | reg_alpha | 2.184 |
| | max_depth | 7 |
| | num_parallel_tree | 9 |
| | min_child_weight | 1 |
| | subsample | 0.621 |
| | learning_rate | 0.948 |
| | gamma | 0.280 |

We used the coefficient of determination ($R^2$), Spearman's rank correlation coefficient (SRCC), root mean squared error (RMSE), and the mean absolute error (MAE) to quantify the performance of developed model.

$$R^2 = 1 - \frac{\sum_{i=1}^{n}(y-\hat{y})^2}{\sum_{i=1}^{n}(y-\bar{y})^2} \quad (1)$$

$$SRCC = 1 - \frac{6\sum d_i^2}{n(n^2-1)} \quad (2)$$

$$MAE = \frac{\sum_{i=1}^{n}|y-\hat{y}|}{n} \quad (3)$$

$$RMSE = \sqrt{\frac{\sum_{i=1}^{n}(y-\hat{y})^2}{n}} \quad (4)$$

where $y$, $\hat{y}$, and $\bar{y}$ denote the true value, predicted value, and the mean of the true values; $d_i$ is the difference in ranks given to the two variables values for each item of the data; n is the number of data points. To explain the importance of each feature, we calculate the SHAP value to analyze the model.

**LLM Fine-tuning**

We fine-tuned three large language models, Llama-3.1-8B-Instruct, Mistral-7B-Instruct-v0.3, and Qwen3-8B, to predict ionic conductivity, using the same training dataset as employed for the GBR model. Fine-tuning improves a general-purpose LLM by training the model on a specific dataset. This helps the model develop a deeper understanding of a particular task, allowing it to produce more relevant responses to queries[50]. Hugging Face Transformers[51] are used to load the pretrained models and employed Unsloth[52] which makes fine-tuning process more efficient even with limited computational resources. Low-rank adaptation (LoRA)[53] was applied to the process, which significantly reduces the number of trainable parameters and GPU memory requirements. Identical fine-tuning hyperparameters were applied to all three models (**Table 2**). The training was conducted using an NVIDIA GeForce RTX 4090 GPU. The batch size per device was set to 4, and the gradient accumulation steps were set to 2 before each backward/update pass. Under these settings, the model training required approximately 10 minutes to complete for 15 epochs. During inference, stochastic decoding was employed, which introduces run-to-run variability in the predicted values. To ensure reproducibility, we repeated the inference five times with different random seeds for each model-case combination. **Table S1** reports the mean and standard deviation of the test MAEs across these runs. The observed standard deviations were less than $\pm 0.08$ log units in all cases, confirming that the stochastic decoding does not significantly affect the reported rankings or conclusions. For each model-case combination, training was conducted at both 10 and 15 epochs, and the epoch with the lower test MAE was selected. The selected model was then evaluated with five random seeds; the main-text metrics correspond to the median run.

**Table 2.** Parameters of LLM models for ionic conductivity prediction model.

| Type | Parameters | Values |
|---|---|---|
| Training | Alpha | 64 |
| | Dropout | 0.05 |
| | Target modules | q_proj, k_proj, v_proj, o_proj, gate_proj, up_proj down_proj |
| | Optimizer | Adam 8bit |
| | Learning Rate | $2 \times 10^{-4}$ |
| | Weight Decay | 0.01 |
| | Scheduler | Cosine |
| | Epochs | 10, 15 |
| Inference | Max new tokens | 12 |
| | Do sample | True |
| | Temperature | 0.7 |

**Table 3.** LLM prompt format.

| LLM Prompt | | |
|---|---|---|
| System prompt | You are a domain expert in materials science specializing in ionic transport. Your task is to predict the logarithm (base-10) of the ionic conductivity of a solid electrolyte. You will be given the material's chemical formula and property description. Output ONLY a single numeric value (the predicted log10(conductivity)) on one line. Do not print units, JSON, labels, extra spaces, or any additional text. Be as precise as possible (use sufficient decimal places) | |
| Input prompt | **Case1**<br>formula: value | **Case 2**<br><formula: value><br><symmetry: value> |
| | **Case 3**<br><formula: value><br><disorder: value> | **Case 4**<br><formula: value><br><symmetry: value><br><disorder: value> |
| Output prompt | value | |

The prompt templates employed for training utilized a zero-shot strategy, as summarized in **Table 3**[54]. Processing solid-state electrolyte (SSE) crystallographic data leads to a complication because of the presence of disorder (i.e., partial occupancy) in the structure files, which is introduced to model the various elemental site-occupancies[55]. Since the standard tool to convert CIFs to text, Robocrystallographer[32], cannot parse CIFs containing such disorder, the chemical formula, symmetry space group name, and atom site occupancy were obtained directly from the CIF files to construct the input prompts. Also, direct use of raw CIF files as text input was avoided because long floating-point atomic coordinates are tokenized into fragmented digit-level subunits, preventing the LLM from interpreting geometric information and introducing substantial numerical noise. Curated structural descriptions preserve essential structural features in an efficient manner, leading to more stable fine-tuning.

Four distinct cases were designed for these inputs: Case 1 included only the chemical formula and served as the baseline for all subsequent cases; Case 2 incorporated the symmetry group; Case 3 added a disorder classification. This classification was categorized in four classes – "Li cation disorder", "Other atom disorder", "Li cation and Other atom disorder" and "No disorder" – specifically to differentiate between disorder in the Li-ion sites and other atoms, given that higher levels of Li-vacancy disorder are known to significantly increase Li-ion conductivity[56]. Finally, Case 4 combined all available information including the chemical formula, symmetry group, and disorder classification.

The output prompt required the LLM to predict the logarithm of the ionic conductivity (without unit). An illustrative example of the LLM input text is provided in **Table 4**. The chemical formula and space group were obtained from the CIF file. Based on the partial atomic occupancies of 0.86 and 0.62 for Li and 0.57 and 0.43 for Y, the disorder was described as exhibiting 'Li cation and Other atom disorder'.

**Table 4.** Example LLM input text from cif file

| CIF File | | | LLMs Input |
|---|---|---|---|
| Chemical Formula | Li3YCl6 | | |
| Symmetry group | P-3m1 | | |
| Atom site occupancy (X, Y, Z, Partial Occupancy (P.O)) | | | |
| Li0 | X: 0.00 | Y: 0.33 | **Case 1)** formula: Li3YCl6 |
|  | Z: 0.50 | P.O: 0.86 | |
| Li1 | X: 0.00 | Y: 0.33 | **Case 2)** |
|  | Z: 0.00 | P.O: 0.62 | <formula: Li3YCl6> <symmetry: P-3m1> |
| Y2 | X: 0.33 | Y: 0.67 | |
|  | Z: 0.49 | P.O: 0.57 | **Case 3)** |
| Y3 | X: 0.33 | Y: 0.67 | <formula: Li3YCl6> <disorder: Li cation and Other atom disorder > |
|  | Z: 0.98 | P.O: 0.43 | |
| Y4 | X: 0.00 | Y: 0.00 | **Case 4)** |
|  | Z: 0.00 | P.O: 1.0 | <formula: Li3YCl6> <symmetry: P-3m1> <disorder: Li cation and Other atom disorder > |
| Cl5 | X: 0.11 | Y: 0.22 | |
|  | Z: 0.78 | P.O: 1.0 | |
| Cl6 | X: 0.13 | Y: 0.56 | |
|  | Z: 0.75 | P.O: 1.0 | |
| Cl7 | X: 0.21 | Y: 0.42 | |
|  | Z: 0.28 | P.O: 1.0 | |

## III. Results

### A. Dataset Composition and Label Distribution (room-temperature dataset)

We first summarize the composition, family coverage, and label distribution of the room-temperature dataset used for model training and evaluation. Because conductivity spans many orders of magnitude, all models are trained and evaluated on $\log_{10}(\sigma/(\text{S cm}^{-1}))$. The final dataset contains 499 labeled crystal structures measured near room temperature (20 – 34 °C, median 27 °C), comprising 321 OBELiX structures, 152 USPEX-generated structures that passed feature extraction, and 26 MP/OQMD structures. Families include NASICON (~20%), garnets (~15%), thiophosphates (~20%), reflecting diversity in the curated SSE dataset but with some biases toward dense oxides/sulfides. The target property $\log_{10}(\sigma/\text{S cm}^{-1})$ spans –17.38 to –1.60, covering 16 orders of magnitude in ionic conductivity. Using $\sigma$ thresholds of $10^{-6}$, $10^{-4}$, and $10^{-3}$ S/cm, the dataset includes 186 low-, 146 moderate-, 103 high-, and 64 superionic-conductivity materials. **Figure 4(a)** illustrates a heat map of elemental occurrence across the dataset. All structures contain Li, while oxygen is present in 349 structures, reflecting the dominance of oxide-type electrolytes in the dataset. Other frequently observed elements include P, S, La, and Ti. **Figure 4(b)** shows the distribution of structures as a function of ionic conductivity. The ionic conductivity range was divided into four categories: low ($< 10^{-6}$ S/cm), moderate ($10^{-6} < \sigma < 10^{-4}$ S/cm), high ($10^{-4} < \sigma < 10^{-3}$ S/cm)[57], and superionic ($> 10^{-3}$ S/cm)[58], the latter being competitive with liquid electrolytes. Because the ionic conductivities in our dataset are distributed over several orders of magnitude, we use log(S/cm) as the target property rather than the raw conductivity σ. Directly training on σ (S/cm) leads to an inherent imbalance in the loss function, such as structures with high conductivity dominate the error.

Distribution of training sets from the diversity selection is shown in **Figure 4(d).** In addition, among the 321 CIF structures in OBELiX dataset, 245 show partial occupancies, as summarized in **Figure 4(c)**. In our workflow, all structures were explicitly labeled as disordered (i.e., containing partial occupancies, PO≠1) or non-disordered (PO=1), and this information was incorporated as a feature for training the GBR model and also fine-tuning the LLM models.

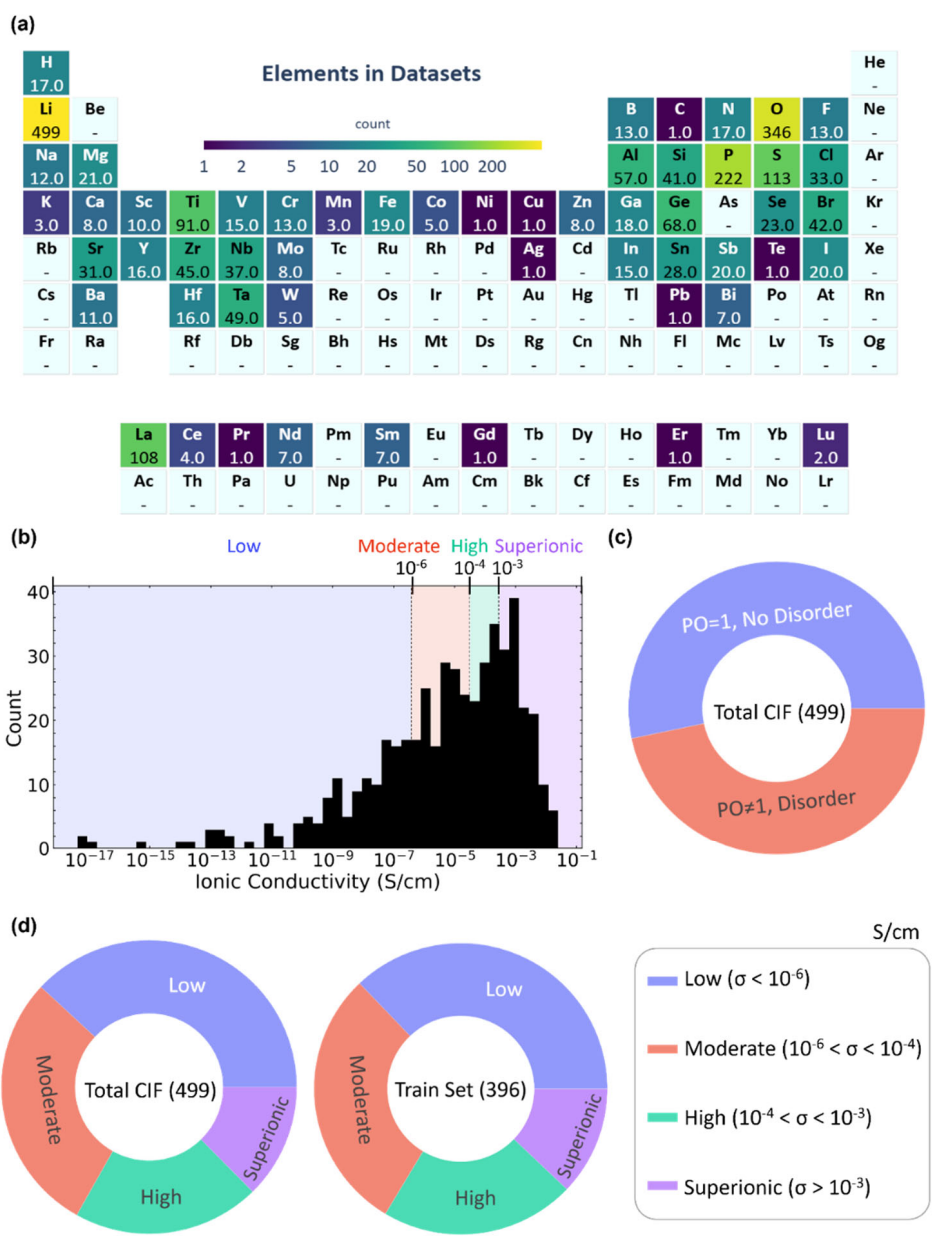

**Figure 4.** (a) The distribution of elements contained in the dataset is shown on the periodic table, generated using pymatviz[59]. (b) Distribution of CIF structures across ionic conductivity ranges, categorized as low (<$10^{-6}$ S/cm), moderate ($10^{-6}$-$10^{-4}$ S/cm), high ($10^{-4}$-$10^{-3}$ S/cm), and superionic (>$10^{-3}$ S/cm); (c) Proportion of structures with partial occupancies among the entire set of CIF files. (d) Ionic conductivity distribution of the dataset used for model training compared with entire dataset.

The dataset collected in this study consists exclusively of inorganic electrolytes. The top 6 families of solid-state electrolytes represented in the overall dataset are summarized in **Table 5**. The full dataset is primarily composed of oxide families (garnet, NASICON, perovskite) and sulfide families (LGPS, argyrodite). Overall, the room-temperature dataset spans many orders of magnitude in conductivity and is dominated by oxide and sulfide chemistries, while also containing a high fraction of partially occupied (i.e., disordered) CIFs. This motivates our two-track modeling strategy: GBR tests whether global geometric descriptors help beyond composition while LLM prompts explicitly encode disorder/symmetry metadata that standard CIF-to-text tools struggle to capture.

**Table 5.** Top six solid-state electrolyte families represented in the overall dataset.

| Rank | Family | Ratio |
| --- | --- | --- |
| 1 | Li-NASICON (Oxide) | 0.21 |
| 2 | Garnet (Oxide) | 0.15 |
| 3 | Perovskite (Oxide) | 0.12 |
| 4 | Argyrodite (Sulfide) | 0.12 |
| 5 | LISICON (Oxide) | 0.08 |
| 6 | Thio-LISICON (Sulfide) | 0.05 |

## B. USPEX+CHGNet Structure Generation and Validation

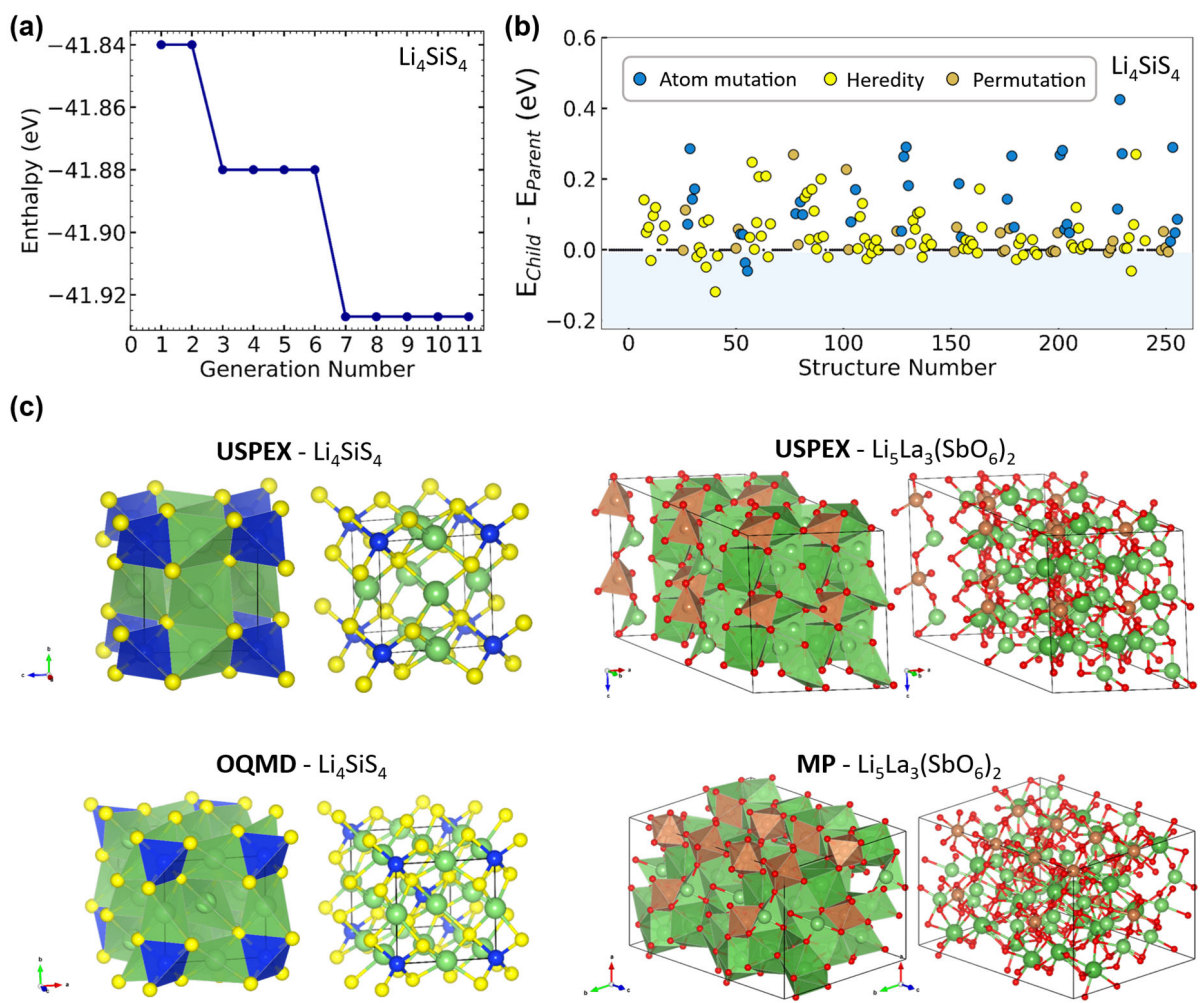

**Figure 5.** (a) Enthalpy change of $Li_4SiS_4$ across generations. (b) Enthalpy variation ($E_{child}$–$E_{parent}$) during structural mutation. (c) Structures generated from compositions with USPEX compared with structures collected from MP and OQMD.

**Figure 5** shows the structure generation results using USPEX. As shown in **Figure 5(a),** for the representative sample structure ($Li_4SiS_4$), the enthalpy generally decreased stepwise with the progress of 11 generation cycles. **Figure 5(b)** displays the entire population of structures generated during this process as a function of enthalpy difference between the child and parent structures ($E_{child}$ - $E_{parent}$) for each variation. Points with a value below 0.0 indicate a successful reduction in enthalpy. Among 3 types of genetic operations (atom mutation, heredity, permutation) at each generation, the heredity operation (both parent structures are cut into two pieces and recombined) leads to enthalpy reductions. The representative sample structures ($Li_4SiS_4$ and $Li_5La_3(SbO_6)_2$) obtained from USPEX were further compared with those reported in the literature, as shown in **Figure 5(c).** To quantitatively assess the similarity between the USPEX-generated structures and reference geometries from the Materials Project and OQMD, we constructed supercells for each structure pair and compared their lattice parameters as well as atomic arrangements (**Table 6**). The supercell lattice lengths showed close agreement across examined cases, indicating that the global geometric scale of generated structures is consistent with MP/OQMD structures. We further computed the root-mean-square displacement (RMSD) of atomic positions using the StructureMatcher module in pymatgen. For $Li_4SiS_4$, the calculated RMSD was 0.320 Å, while for $Li_5La_3(SbO_6)_2$, the RMSD was 0.521 Å. These RMSD values (0.320 – 0.521 Å) represent approximately 12 – 20% of typical Li-S bond lengths (~2.4 – 2.6 Å), confirming that positional deviations remain within chemically reasonable limits.

**Table 6.** Structural similarity between USPEX-generated structure and MP/OQMD-collected structures.

|  | $Li_4SiS_4$ (USPEX) | $Li_4SiS_4$ (OQMD) | $Li_5La_3(SbO_6)_2$ (USPEX) | $Li_5La_3(SbO_6)_2$ (MP) |
|---|---|---|---|---|
| Supercell atom # | 18 | 18 | 88 | 88 |
| _cell_length_a | 11.156 | 11.587 | 11.976 | 11.199 |
| _cell_length_b | 11.156 | 11.587 | 10.332 | 11.213 |
| _cell_length_c | 11.158 | 10.376 | 11.976 | 11.218 |
| RMSD (Å) | 0.320 | | 0.521 | |

When visualizing $Li_4SiS_4$ structure collected from the OQMD, we observe similarity in the atomic arrangements. Li atoms form tetrahedrons with four neighboring S atoms, and Si atoms also form tetrahedrons with four neighboring S atoms. However, structures where S atoms form hexahedrons are not present in the USPEX-generated structures. Similarly, for $Li_5La_3(SbO_6)_2$, the Li atoms forming octahedrons with six neighboring O atoms are consistent with structure from MP. However, the Sb atoms, which form octahedrons in the MP structure, appeared as hexahedron in the USPEX-generated structure. Despite these differences, the polyhedron structures of Li atoms in both USPEX-generated sample structures are identical to those in the DFT optimized structures. While some atoms exhibited different polyhedron structures, the elements of the adjacent atoms are consistent between both structures, indicating a structural similarity with optimized structures and those from MP and OQMD database.

To assess the physical plausibility of the 152 crystal structures generated by USPEX and subsequently relaxed using CHGNet, we performed validation using automated analysis tools implemented in pymatgen. We evaluated the minimum interatomic distances across all structures to identify any unphysically short bonds that would indicate structural instability or unrealistic atomic configurations. Except for one structure, all minimum bond lengths exceeded 1.3 Å, which falls within expected chemically reasonable ranges for the elements considered. The single outlier featured an O-H bond of 0.96 Å, which is consistent with experimental hydroxyl bond lengths (~0.97 Å), confirming it is physically valid. Furthermore, to assess whether the USPEX-predicted structures correspond to stable crystal phases, each optimized structure was further equilibrated for an additional 20 ps NPT MD simulation at 300 K and 1 atm using CHGNet. **Figure S1** shows the potential energy of all 152 structures during the MD trajectory. Throughout the simulations, no atom loss or structural collapse was observed. These results suggest that the generated structures maintain structural integrity under experimentally measured conditions on the timescale of the simulation. These validations indicate that the USPEX with CHGNet workflow produces physically plausible geometries that are broadly consistent with reference database structures and remain stable during short finite-temperature MD. The resulting generated structures expand structural coverage for compositions lacking experimentally resolved CIFs, enabling structure-aware model training at room temperature.

## C. Gradient-boosted Trees: Effect of Compositional versus Geometric Descriptors and SHAP Interpretation

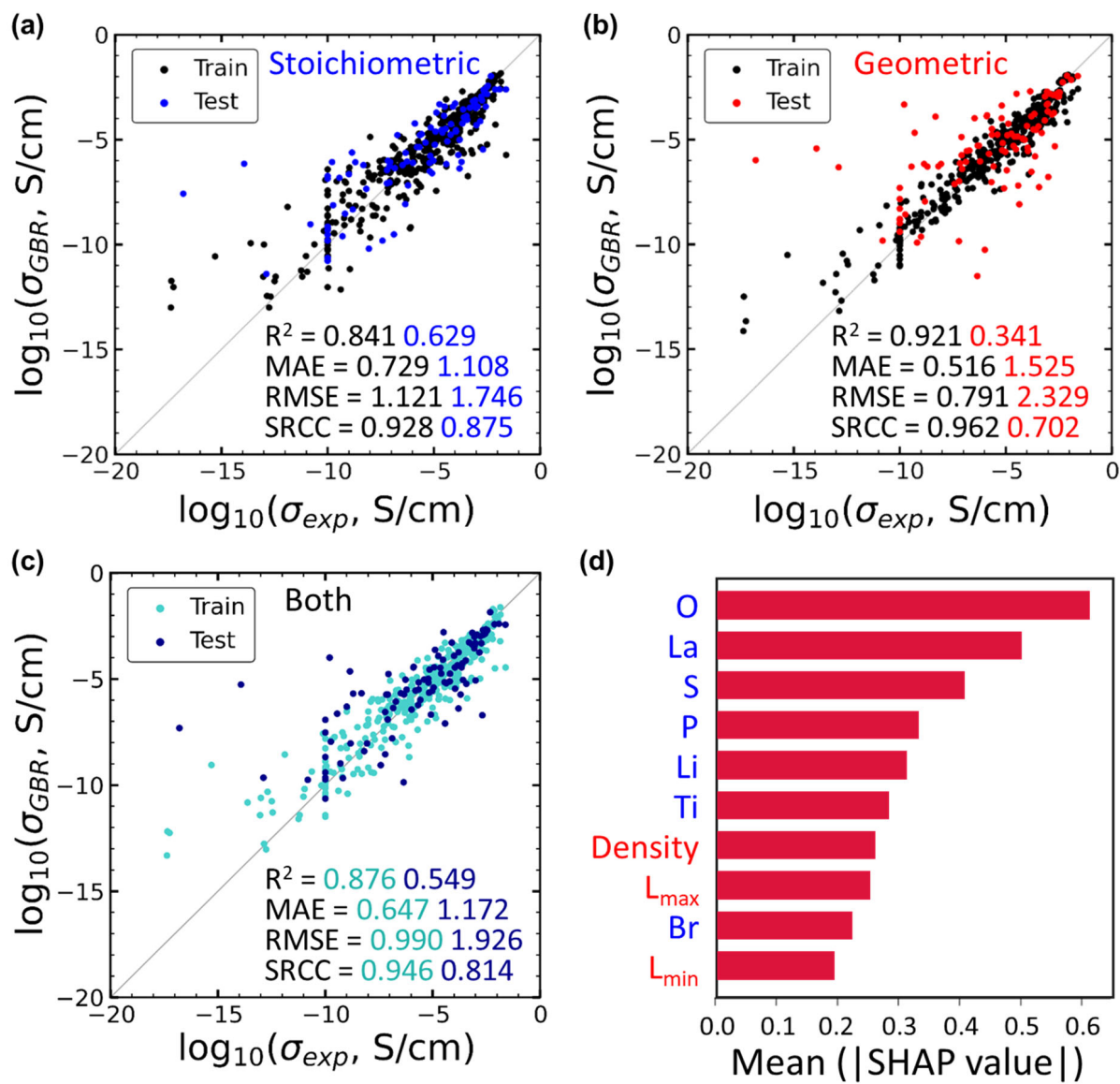

**Figure 6.** (a) Stoichiometric features only. (b) Geometric features only. (c) All features. (d) SHAP analysis of feature importance using all features and geometric features.

The performance of the Gradient Boosting Regressor (GBR) model was evaluated using stoichiometric features, geometric features, and a combination of both. **Figures 6(a)-(c)** visualize the predictive performance of the GBR model when using the stoichiometric, geometric, and combined features, respectively. **Figure 6(d)** presents the results of a feature importance analysis for the model trained with combined features, quantified by SHAP values. When trained solely on stoichiometric features, the model achieved a test MAE of 1.108 and SRCC of 0.875, indicating strong ranking capability. Geometric features alone yielded the lowest accuracy (MAE = 1.525, SRCC = 0.702). Combining both feature types gave a test MAE of 1.172, and SRCC of 0.814. Although the combined model did not surpass the stoichiometric-only model in either test MAE or SRCC, the higher training $R^2$ (0.876 vs. 0.841) alongside a lower test $R^2$ (0.549 vs 0.629) indicates that the additional geometric degrees of freedom introduce a degree of overfitting at the current dataset size (n = 499). This gap may narrow as larger structure-labeled conductivity datasets become available.

The SHAP analysis (**Figure 6d**) confirms that stoichiometric descriptors dominate the combined model, occupying seven of the top ten positions. The oxygen ratio was ranked first by a substantial margin, consistent with the prevalence of oxide electrolytes (~60% of the dataset), where oxygen simultaneously stabilizes the lattice and modulates Li+ conduction bottlenecks[60]. The next most important features, La, S, P, Li, Ti, and Br, reflect the dominance of garnet, NASICON, and sulfide families in the training data. Among geometric descriptors, Density (7th overall), $L_{max}$ and $L_{min}$ enter the top ten. The reordered lattice parameters provide axis-invariant measures of unit cell dimensions, avoiding the labeling bias of conventional a/b/c assignments that is particularly relevant for the ~39% of structures with P1 symmetry. This SHAP ranking reflects correlations within this dataset rather than universal causal relationships. Therefore, incorporating richer geometric descriptors, such as pore-connectivity metrics or channel topology, may further improve the model generalizability in future work.

## D. LLM Fine-Tuning: Prompt Ablation with CIF-derived Metadata

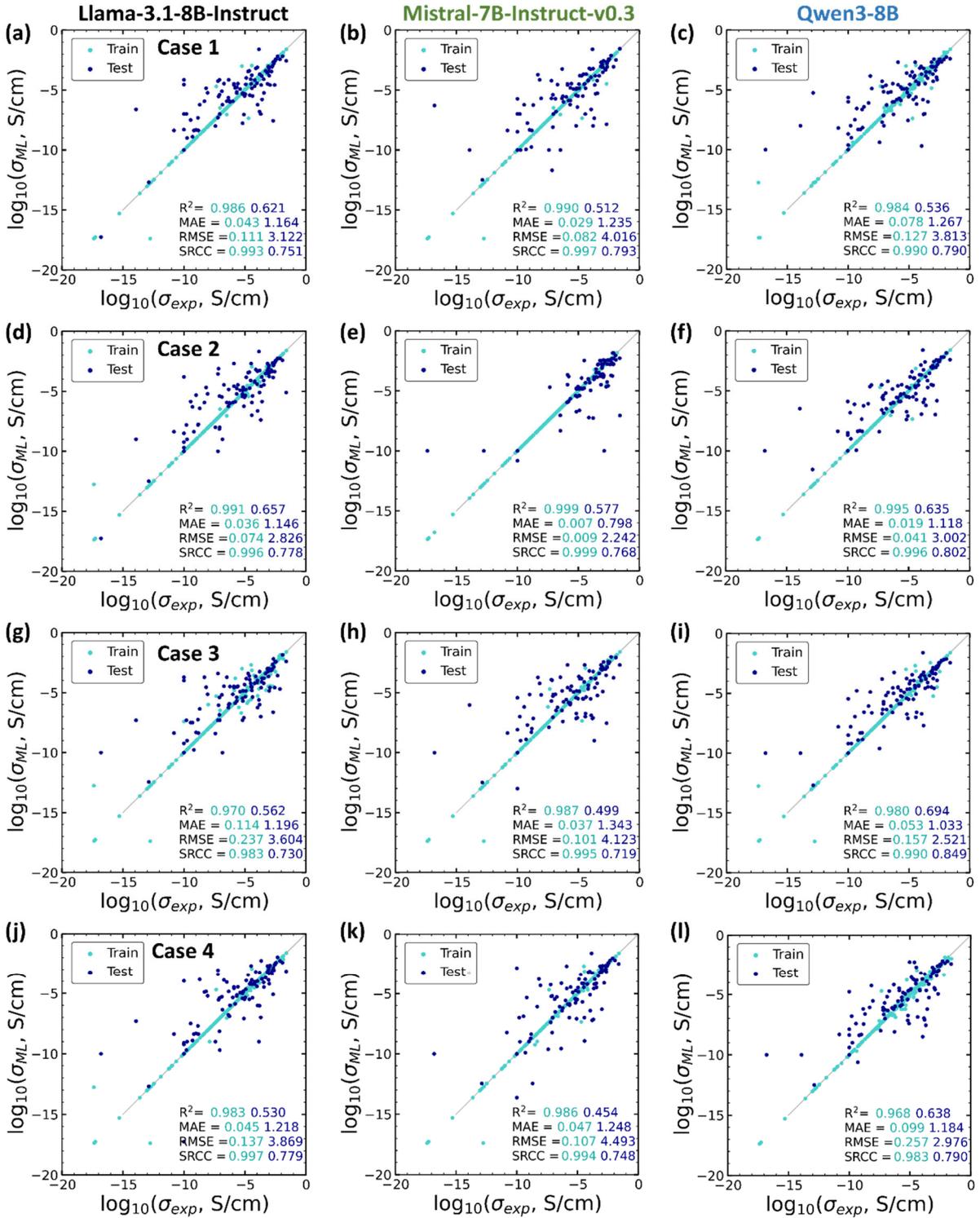

**Figure 7.** (a)-(c) The results for Case 1. (d)-(f) The results for Case 2. (g)-(i) The results for Case 3. (j)-(l) The results for Case 4.

Having established the importance of structural features using GBR, we next evaluated whether

Large Language Models could capture these structure-property relationships directly from textual descriptors, thereby bypassing the need for explicit geometric feature engineering. **Figure 7** illustrates the prediction results for the train and test sets across four different cases of three Large Language Models (LLMs): Llama-3.1-8B-Instruct, Mistral-7B-Instruct-v0.3, and Qwen3-8B, where training was conducted at both 10 and 15 epochs for each configuration and the results with the lower test MAE are reported. The performance is evaluated by plotting the logarithm of LLM-predicted conductivity against the logarithm of the experimental ionic conductivity at room temperature. For evaluation metrics, $R^2$, MAE, RMSE, and SRCC are used to quantify predictive accuracy on the test set. Among the three models, Mistral-7B-Instruct-v0.3 demonstrates consistently strong predictive performance, particularly in Case 2 **(Figure 7(e))**. In this case, which incorporates compositional features and symmetry information, the model achieves the lowest MAE of 0.798 and RMSE of 2.242 across all models and feature cases. Qwen3-8B achieves its best performance in Case 3 **(Figure 7(i))**, where compositional information and disorder descriptors are provided as input. In this setting, it records the highest SRCC of 0.849 among all evaluated configurations, indicating the strongest rank-ordering capability. Across all three models, Case 2, which incorporates symmetry information alongside the chemical formula, consistently achieves lower test MAE than Case 1, which relies solely on the chemical formula (**Figure 7(a–f)**), indicating that symmetry metadata provides meaningful additional information for ionic conductivity prediction beyond composition alone. However, when all available features are combined in Case 4 (**Figure 7(j–l)**), the test MAE increases relative to each model's best-performing case. This suggests that including all metadata simultaneously may introduce redundant or conflicting information, leading to overfitting or noise in the fine-tuning process. One possible explanation is that when symmetry and disorder information are simultaneously provided, the model may learn spurious correlations between redundant descriptors, for instance, certain space groups are inherently associated with specific disorder patterns, creating collinear features that disrupt generalization. These results highlight that selective use of structural metadata, rather than exhaustive inclusion, is more effective for LLM-based conductivity prediction. The optimal metadata combination is model-dependent, suggesting that prompt content interacts differently with each model's pretraining representation.

**E. Cross-Model Comparison and Applicability**

To place our results in context, **Table 7** compares our room-temperature (RT) models with the OBELiX benchmark models trained on experimental RT conductivity data. Because the OBELiX benchmark evaluated composition-based models (RF, MLP) on both the full test set (n = 121) and the CIF-only test subset (n = 67), we report their CIF-only test MAE for a more consistent comparison with our structure-labeled models. On our curated RT dataset ($n_{train}$ = 396, $n_{test}$ = 103), the fine-tuned LLMs achieved competitive accuracy: Mistral-7B (Case 2: formula and symmetry) achieves the lowest test MAE of 0.798, while Qwen3-8B (Case 3: formula and disorder) achieves the highest SRCC of 0.849 (MAE = 1.033), while the best GBR model using stoichiometric features alone achieves MAE = 1.108. In log10 units, an MAE of 1.033 corresponds to an average multiplicative uncertainty of ~10.7x in conductivity. For reference, OBELiX reports an experimental reproducibility floor of MAD = 0.41 log units from repeated measurement of identical compositions,[20] indicating that all current models remain well above the intrinsic experimental uncertainty. Relative to prior experimental-data-driven models, our controlled RT evaluation demonstrates that both approaches achieve competitive accuracy, with the GBR model offering interpretability through SHAP analysis and the LLM providing a low-preprocessing alternative that bypasses explicit feature engineering. We note that a direct numerical comparison between the OBELiX benchmark models and ours should be interpreted with caution: the two evaluations differ in dataset size (599 vs. 499), test set comparison (67 vs. 103 structures), feature space, and splitting strategy. In particular, our dataset includes 152 USPEX-generated structures not present in OBELiX, which may shift the conductivity distribution. The comparison is therefore intended to place our results in an approximate context rather than claim performance advantages.

Table 7. Test set MAE and SRCC for experimental-data-driven ionic conductivity models. MAE is reported in $\log_{10}(\sigma/(\text{S cm}^{-1}))$ units.

| Model | Features | Number of Train Data | Number of Test Data | Test MAE ($\log_{10}(\sigma/(\text{S cm}^{-1}))$) | SRCC | Ref. |
|---|---|---|---|---|---|---|
| RF[a] | Composition | 478 | 67 | 1.59[b] | - | [20] |
| MLP[a] | Composition, space group, lattice | 478 | 67 | 1.72[b] | - | [20] |
| PaiNN[a] | Crystal structure (CIF) | 254 | 67 | 2.60 | - | [20] |
| GBR | Stoichiometric | 396 | 103 | 1.108 | 0.875 | This work |
| GBR | Stoichiometric + Geometric | 396 | 103 | 1.172 | 0.814 | This work |
| Mistral-7B | Text prompt (formula, symmetry) | 396 | 103 | 0.798 | 0.768 | This work |
| Qwen3-8B | Text prompt (formula, disorder) | 396 | 103 | 1.033 | 0.849 | This work |

[a]OBELiX benchmark models. RF and MLP were trained on the full OBELiX training set (n = 478) using composition, space group, and lattice parameters. PaiNN was trained on the CIF subset (n = 254). [b]Although RF and MLP were trained on the full set, the reported MAE corresponds to the CIF-only test subset (n = 67) to enable comparison with geometric models. The full-set test MAE (n = 121) is lower (~1.06 for RF); see Therrien et al.[20] for details.

To assess model performance across different families, test set MAE values were compared by category as shown in **Table 8**. Among sulfide electrolytes, argyrodite structures containing halide anions exhibit distinct conductivity ranges compared to conventional non-halide sulfides and oxides[61]. Therefore, we classified the solid electrolytes into oxides, non-halide sulfides, halides, and others, and compared the prediction performance for each category.

**Table 8.** Test set model performance by electrolyte family. MAE is reported in $\log_{10}(\sigma/(\text{S cm}^{-1}))$ units. GBR uses stoichiometric features only; LLM results are from fine-tuned Qwen3-8B (Case 3: formula + disorder).

| Family | Representative Types | $n_{test}$ | GBR MAE | GBR SRCC | LLM MAE | LLM SRCC |
|---|---|---|---|---|---|---|
| Oxide | Garnet, NASICON, Perovskite | 66 | 0.90 | 0.88 | 0.94 | 0.80 |
| Sulfide (non-halides) | LGPS, thio-phosphate, thio-LISICON | 9 | 1.15 | - | 0.72 | - |
| Sulfide (halides) | Argyrodite ($Li_6PS_5X$, X = Cl, Br, I) | 15 | 0.51 | 0.84 | 0.50 | 0.88 |
| Others (non-S /O) | Halides, Phosphates | 13 | 2.80 | - | 2.35 | - |
| All | | 103 | 1.108 | 0.875 | 1.033 | 0.849 |

**Table 8** presents the per-family test MAE for the GBR (stoichiometric features) and fine-tuned Qwen3-8B (Case 3: formula + disorder) models. Sulfide-based electrolytes were predicted with the highest accuracy: the Qwen3-8B model achieves MAE of 0.50 and 0.72 for halide and non-halide sulfides, respectively, while the GBR model achieves 0.51 and 1.15. Oxides, the most well-represented family (n = 66), show reliable predictions from both models (MAE = 0.90 for GBR, 0.94 for Qwen3-8B). The Others category (halides and phosphates, n = 13) shows the lowest accuracy (MAE = 2.80 for GBR, 2.35 for Qwen3-8B), likely due to limited sample size and chemical diversity. Although the GBR model slightly outperforms the LLM for oxides, the LLM achieves lower or comparable MAE across all other families, highlighting its ability to generalize across chemically distinct SSE classes. In addition to MAE, both models show strong rank-ordering capability (SRCC), indicating the usefulness of the developed models when integrated into a computational screening workflow.

Predicted ionic conductivities for representative electrolytes from each family were compared with experimental measurements in **Table 9**. For the sulfide halide family, both models closely reproduce experimentally reported values for argyrodite compounds (Li$_6$PS$_5$Br and Li$_6$PS$_5$I), with deviations of less than one order of magnitude. Non-halide sulfides, including LGPS (Li$_{10}$GeP$_2$S$_{12}$) and Li$_3$SbS$_3$, are also predicted with reasonable accuracy, although the GBR model shows a larger error for Li$_3$SbS$_3$. In the oxide family, the NASICON-type LiGe$_2$(PO$_4$)$_3$ is well predicted by both models, whereas the garnet-type Li$_5$Nd$_3$Sb$_2$O$_{12}$ is substantially overestimated by the GBR model, while the LLM provides a closer prediction. For the Others category, both models show large deviations from experimental values, as exemplified by the pyrophosphate Li$_2$BaP$_2$O$_7$ and the imide Li$_2$Ca(NH)$_2$, where predicted conductivities differ from the measured values by several orders of magnitude. These findings suggest that expanding the experimental dataset to cover a wider range of elements and stoichiometries is essential for improving the generalization of both models. Overall, well-established families such as argyrodites and NASICON-type oxides are predicted accurately, whereas uncommon chemistries expose the extrapolation limits of both approaches.

**Table 9.** Predicted ionic conductivities of representative solid electrolytes from each family compared with experimental values.

| Family | Composition | Exp. log10($\sigma$/S cm$^{-1}$) | GBR log10($\sigma$/S cm$^{-1}$) | LLM log10($\sigma$/S cm$^{-1}$) |
|---|---|---|---|---|
| Oxide | LiGe$_2$(PO$_4$)$_3$ | −6.48 | −6.22 | −6.07 |
| | Li$_5$Nd$_3$Sb$_2$O$_{12}$ | −6.89 | −9.52 | −5.18 |
| Sulfide (non-halides) | Li$_{10}$GeP$_2$S$_{12}$ | −1.61 | −2.61 | −2.45 |
| | Li$_3$SbS$_3$ | −9.00 | −6.65 | −8.32 |
| Sulfide (halides) | Li$_6$PS$_5$Br | −2.56 | −3.42 | −2.64 |
| | Li$_6$PS$_5$I | −3.66 | −4.23 | −4.58 |
| Others (non- S /O) | Li$_2$BaP$_2$O$_7$ | −16.80 | −7.58 | −10.00 |
| | Li$_2$Ca(NH)$_2$ | −5.19 | −6.51 | −3.77 |

## IV. Conclusion

In this work, we developed two complementary data-driven predictors for room-temperature ionic conductivity of solid-state electrolytes, trained on a curated dataset of 499 CIF structures. A GBR model using stoichiometric features alone achieved a test MAE of 1.108 log(S/cm), while SHAP analysis of the combined model revealed that compositional descriptors dominate but geometric features (density, $L_{max}$, $L_{min}$) provide complementary structural information. Fine-tuned LLMs using text prompts derived from CIF metadata achieved competitive accuracy without explicit feature engineering; notably, Mistral-7B achieved the lowest MAE of 0.798 log(S/cm) using formula-and-symmetry prompts, while Qwen3-8B with formula-and-disorder prompts achieved the strongest ranking performance (SRCC = 0.849, MAE = 1.033 log(S/cm)), demonstrating that encoding crystallographic disorder as categorical text descriptors enables LLMs to capture structure-property relationships inaccessible through formula-only prompts.

Overall, our results demonstrate that incorporating structural information into both GBR and LLM models enhances the predictive capability for ionic conductivity. However, the limited range of experimentally measured ionic conductivities currently available constrains a rigorous evaluation of model extrapolation to novel materials. A key limitation of our framework is its reduced extrapolation capability for materials that fall outside the compositional and structural domain of the training set. Because the dataset primarily consists of dense oxide- and sulfide-based solid electrolytes, the model may underperform for systems with highly porous frameworks or well-defined ion-channel geometries whose pore volumes, densities, or channel topologies deviate from those represented in the training structures. Additionally, our computational framework does not explicitly account for the potential energy landscape governing ion migration or the ion–ion correlation effects that can arise at high carrier concentrations. Recent developments in interaction-aware neural potentials, such as the methodology proposed by Gustafsson et al.[62], demonstrate that capturing many-body ion–ion correlations and mapping energy landscapes can provide deeper mechanistic insight into solid-state ionic transport. Incorporating such energy-landscape-based or interaction-aware descriptors lies beyond the scope of the present work but represents an important direction for next-generation solid-state electrolyte design and discovery. Despite these limitations, the competitive accuracy achieved by both GBR and LLM approaches on a modestly sized dataset underscores the promise of data-driven methods for accelerating solid-state electrolyte discovery.

**Supplementary material**

**ACKNOWLEDGMENTS**

This work was supported by the National Research Foundation of Korea (NRF) from a grant funded by the Korea government (MSIT) (RS-2024-00449431). The computational resources were provided by KISTI (KSC-2024-CRE-0412). This research was supported by the Regional Innovation System & Education (RISE) program through the Institute of Regional Innovation System & Education in Busan Metropolitan City, funded by the Ministry of Education (MOE) and Busan Metropolitan City, Republic of Korea (2025-RISE-02-001-020).

**AUTHOR DECLARATIONS**

**Conflict of Interest**

The authors have no conflicts to disclose.

**Author Contributions**

**Haewon Kim:** Data curation; Software; Formal analysis; Investigation; Methodology; Validation; Visualization; Writing – original draft; Writing – review & editing. **Taekgi Lee:** Investigation; Software; Methodology; Validation; Visualization; Writing – original draft; Writing – review & editing. **Seong Eun Hong:** Data curation; Methodology; Investigation. **Kyeong-Ho Kim:** Writing – review & editing. **Yongchul G. Chung:** Conceptualization; Formal analysis; Resources; Supervision; Project administration; Funding acquisition; Writing – review & editing.

## DATA AVAILABILITY

Developed machine learning models and structure generation data are available on https://github.com/Chung-Research-Group/reproducible-workflows/tree/master/2025-SSE, the dataset can be accessed from https://doi.org/10.5281/zenodo.17157647.